\documentclass[a4paper,11pt]{article}
\usepackage{jheppub}
\usepackage[english]{babel}
\usepackage{bigints}
\usepackage{amsmath}
\usepackage{amssymb}
\usepackage{amsthm}
\usepackage{graphicx}
\usepackage{caption}
\usepackage{subcaption}
\usepackage{tikz}
\usepackage{xcolor}
\newcommand{\plotlegend}[3]{%
  \begin{tikzpicture}[baseline, scale=0.9] 
    \node[
      draw,
      rounded corners=1.5pt,
      line width=0.5pt,
      inner sep=2pt,
      fill=white
    ] {
      \begin{tabular}{@{}l@{}}
        \raisebox{0.2ex}{\tikz{\draw[plotcolorone, thick] (0,0)--(0.07,0);}}~~#1\\[1pt]
        \raisebox{0.2ex}{\tikz{\draw[plotcolortwo, thick] (0,0)--(0.07,0);}}~~#2\\[1pt]
        \raisebox{0.2ex}{\tikz{\draw[plotcolorthree, thick] (0,0)--(0.07,0);}}~~#3
      \end{tabular}
    };
  \end{tikzpicture}%
}
\definecolor{plotcolorone}{RGB}{31,119,180}   
\definecolor{plotcolortwo}{RGB}{255,127,14}   
\definecolor{plotcolorthree}{RGB}{44,160,44}  
\usetikzlibrary{arrows.meta, decorations.pathmorphing, calc, intersections}
\newcommand*\circled[1]{\tikz[baseline=(char.base)]{
            \node[shape=circle,draw,inner sep=2pt] (char) {#1};}}

\DeclareMathAlphabet{\mathdutchcal}{U}{dutchcal}{m}{n}

\title{Ramp and plateau in bulk correlators within the disk topology in JT gravity}
\author{Cristiano Germani}\author{and Mickael Komendyak} 
\affiliation{Departament de F\'{i}sica Qu\`{a}ntica i Astrof\'{i}sica and Institut de Ci\`{e}ncies del Cosmos,\\ Universitat de Barcelona, Mart\'{i} i Franqu\`{e}s 1, 08028 Barcelona, Spain}
\emailAdd{germani@icc.ub.edu}\emailAdd{komendyak@icc.ub.edu}
\date{}
\abstract{We study bulk two-point correlation functions of a massless scalar field in Jackiw–Teitelboim gravity around the eternal black hole saddle. While same-side correlators exhibit exponential decay, two sided correlators, at the next to leading order in steepest descent approximation, exhibit a ramp followed by a plateau after the initial semi-classical exponential decay. Our results indicate that the late-time saturation of two-sided correlators is already visible within the perturbative saddle expansion of the bulk path integral, without invoking nonperturbative $e^{-S}$ effects. Finally, we show that the dip-time, defined as the minimum of the correlator, grows inversely with the black hole temperature, consistent with expectations from the holographic dual.}

\begin{document}
\maketitle

\newpage


\section{Introduction}

Black holes are known to radiate. Hawking’s seminal computation \cite{hawking1974, hawking1975} shows that an evaporating black hole emits an almost thermal spectrum characterized only by its macroscopic parameters (mass, charge, angular momentum). If this conclusion is extrapolated to the endpoint of evaporation, one is led to the conclusion that pure states evolve into mixed states, apparently violating unitarity and threatening the foundations of quantum mechanics. This conflict is commonly referred to as the black hole information paradox \cite{hawking1976, page1993_bh}.

One might hope that the missing information is merely hidden behind the event horizon, inaccessible to an exterior observer until the very final stages of evaporation. Page famously argued, however, that if black hole evaporation is unitary, the entanglement entropy of the Hawking radiation must begin to decrease long before the black hole shrinks to Planckian size \cite{page1993_bh,page1993_av}. The corresponding Page time typically occurs when roughly half of the initial Bekenstein–Hawking entropy has been radiated away, at which point the black hole is still semiclassical and large. The unavoidable conclusion is that the standard Hawking calculation — firmly rooted in semiclassical gravity — must already miss an essential ingredient well before curvatures become extreme.

Confronted with this puzzle, it is natural to seek gravitational models where quantum effects are tractable and information flow can be analyzed exactly. Jackiw–Teitelboim (JT) \cite{jackiw1985,teitelboim1983} gravity — two-dimensional dilaton gravity with a negative cosmological constant — has emerged as a particularly fruitful setting. Classically, it is simple: the metric has constant negative curvature and the dilaton carries the nontrivial gravitational degree of freedom. Quantum mechanically, it is nontrivial: the theory admits black holes with finite entropy and a well-defined Hamiltonian description of boundary dynamics governed by the Schwarzian action \cite{Maldacena:2016upp,almheiri2015jt, engelsoy2016}. Moreover, semiclassical JT gravity is conjectured to be the gravitational dual of the universal low-energy limit of Sachdev–Ye–Kitaev (SYK)-like models (from now on simply SYK theory) \cite{sachdev2010,cotler2017blackholes}. This theory corresponds to a large number of interacting Majorana fermions at low energies (with respect to the coupling constants), see e.g. \cite{maldacenastanford2016}.

In SYK, disorder-averaged correlators exhibit the characteristic dip–ramp–plateau structure associated with chaotic quantum systems \cite{cotler2017blackholes,saad2018semiramp}. These features are most prominently discussed in the context of spectral observables and ensemble-averaged quantities, where nonperturbative effects related to level discreteness play a central role. By contrast, the two-sided correlators in the thermofield-double (TFD) state of two entangled SYK systems, which is the boundary dual of bulk observables connecting the two asymptotic regions of the eternal black hole, have received far less attention. This raises the question of whether the late-time behavior of such TFD correlators necessarily relies on nonperturbative spectral effects.

In the bulk JT-gravity description, recent efforts to reproduce the ramp in the boundary correlators incorporate connected wormhole topologies into the gravitational path integral \cite{saad2018semiramp,saad2019jt,saad2020nonpert,stanfordwitten2020,mertens2023review,blommaert2019}. The leading connected contribution arises from the so-called double-trumpet Euclidean wormhole \cite{mertens2023review}, which yields the linear ramp, while the plateau is commonly attributed to genuinely non-perturbative effects in Newton’s constant \cite{marolfmaxfield2020,saadyang2021half}. In these approaches, the gravitational path integral is supplemented by a topological term that governs the relative weighting of geometries and enables the inclusion of connected wormhole contributions. While such a sum over topologies can be implemented at the boundary, its precise interpretation for bulk correlators remains subtle \footnote{CG thanks Vijay Balasubramanian for pointing this out.}. From the viewpoint of the steepest-descent expansion, one may instead ask whether the relevant saddle structure is already encoded in the original gravitational action, without the need to introduce additional topological weighting.

Following the philosophy of \cite{germani}, in contrast to previous approaches, we work entirely within the disk topology for bulk correlators and show that while the same side (LL) correlator exhibit the typical semicalssical decay,  a dip–ramp–plateau structure arises already from the perturbative expansion of the correlation function evaluated across two maximally entangled black hole exteriors. At leading order, the two sided (LR) correlator exhibits pure exponential decay. However, at the next perturbative order in the gravity coupling constant, subleading corrections generate a linear ramp that ultimately saturates at a constant, power-suppressed plateau. This shows that non-decaying late-time correlations are already encoded within perturbative fluctuations around the principal Lorentzian saddle of the gravitational path integral, without invoking additional topologies or genuinely nonperturbative $e^{-S}$ effects.

\section{JT Gravity}

Jackiw–Teitelboim gravity is a two-dimensional model of dilaton gravity \cite{jackiw1985,teitelboim1983,mertens2023review} defined by a bulk action involving the product of the dilaton scalar field $\Phi$ and the Ricci scalar $R$, supplemented by the Gibbons-Hawking-York (GHY) boundary term. To focus on 2d Anti-de Sitter space (AdS$_2$), we set $\Lambda=-\frac{1}{L^2}$ and work in units where the AdS$_2$ length is set to $L=1$:

\begin{equation}
\label{JT_Action}
S_{\mathrm{JT}}[g,\Phi]
   = \frac{1}{16\pi G_N}\left(\int_{\mathcal{M}}\Phi(R+2)\sqrt{-g}\, dx^2
     \,+\,2\oint_{\partial\mathcal{M}}\Phi_{|\partial}\,(K-1)\sqrt{-h}\, dx\right) .
\end{equation}


\subsection{Metric Solutions}

In two dimensions, the Ricci scalar $R$ fully determines the local geometry since the Riemann tensor has only one component. Thus, varying eq.(\ref{JT_Action}) with respect to $\Phi$ fixes the background manifold to be locally AdS$_2$ through $R=-2$. This determines the metric up to coordinate transformations.\\

In global coordinates, the AdS$_2$ metric takes the form:

\begin{equation}
\label{Global patch}
    ds^2=\frac{1}{\cos^2(z_G)}(-dt_G^2+dz_G^2)=-\frac{1}{\cos^2(\frac{1}{2}(u_G-v_G))}du_Gdv_G,
\end{equation}

\noindent where $z_G\in[-\frac{\pi}{2},\frac{\pi}{2}]$, $t_G\in\mathbb{R}$, and the lightcone coordinates are defined as $u_G=t_G+z_G$ and $v_G=t_G-z_G$. These coordinates cover the maximal extension of the AdS$_2$ manifold. \\

One can get to the Poincaré patch through the transformations $u_G=2\tan^{-1}(u_P)-\frac{\pi}{2}$, and $v_G=2\tan^{-1}(v_P)+\frac{\pi}{2}$:

\begin{equation}
\label{Poincaré patch}
    ds^2=\frac{1}{z_P^2}(-dt_P^2+dz_P^2)=-\frac{4}{(u_P-v_P)^2}du_Pdv_P,
\end{equation}

\noindent where $z_P\in(0,\infty)$, and $t_P\in\mathbb{R}$. This patch covers only a region inside the global manifold. The Poincaré time $t_P$ does not correspond to the global time $t_G$ and therefore one expects the Poincaré vacuum to be different from the global vacuum. However, as was shown in \cite{spradlin1999vacuum,danielsson1998} this is not the case and both vacua are in fact equivalent. \\

\begin{figure}[ht]
  \centering
  \begin{subfigure}[t]{0.45\textwidth}
    \centering
    \begin{tikzpicture}[>=Stealth, font=\small]
  \def\Tmax{4}    
  \def\Zmax{3}    

  \fill[blue!20] (0,3) -- (\Zmax,0) -- (0,-3) -- cycle;

  \draw[thick] (0,-\Tmax) -- (0,\Tmax)  ;
  \draw[thick] (3,-\Tmax) -- (3,\Tmax)  ;

  \draw[dashed] (0,3) -- (\Zmax,0)
  node[midway,sloped,above] {$z_P=\infty$};
  \draw[dashed] (0,-3) -- (\Zmax,0)
  node[midway,sloped,below] {$z_P=\infty$};
  \draw[->,thick]
    (0.5,-1.8) 
      to[out=45,in=-45]        
    (0.5, 1.8) 
    node[sloped,above] {$t_P$};

  \draw[->, thick] (0,0) -- (4,0) node[right] {$z_G$};
  \draw[->, thick] (0,-\Tmax) -- (0,\Tmax) node[above] {$t_G$};
 
  \node at (1.5,0.2) {Poincar\'e Patch};
  \node at (1.5,3.5) {Global Patch};

\end{tikzpicture}
    \caption{}
  \end{subfigure}\quad
  \begin{subfigure}[t]{0.45\textwidth}
    \centering
    \begin{tikzpicture}[>=Stealth, font=\small]
  \def\Tmax{2}    
  \def\Zmax{2}    

  \fill[blue!20] (-2,2) -- (0,0) -- (-2,-2) -- cycle;
  \fill[blue!5] (-2,2) -- (0,0) -- (2,2) -- cycle;
  \fill[blue!20] (2,2) -- (0,0) -- (2,-2) -- cycle;
  \fill[blue!5] (-2,-2) -- (0,0) -- (2,-2) -- cycle;

  \draw[thick] (-\Zmax,-4) -- (-\Zmax,4)  ;
  \draw[thick] (-\Zmax,2) -- (\Zmax,2)  ;
  \draw[thick] (\Zmax,4) -- (\Zmax,-4)  ;
  \draw[thick] (\Zmax,-\Tmax) -- (-\Zmax,-\Tmax)  ;

  \draw[dashed] (-\Zmax,-\Tmax) -- (0,0)
  node[midway,sloped,below] {$z^I_R=\infty$};
  \draw[dashed] (\Zmax,\Tmax) -- (0,0)
  node[midway,sloped,above] {$z^{II}_R=\infty$};
  \draw[dashed] (-\Zmax,\Tmax) -- (0,0)
  node[midway,sloped,above] {$z^I_R=\infty$};
  \draw[dashed] (\Zmax,-\Tmax) -- (0,0)
  node[midway,sloped,below] {$z^{II}_R=\infty$};
  \draw[->,thick]
    (-\Zmax+0.2,-\Tmax+0.4) 
      to[out=45,in=-45]       
    (-\Zmax+0.2, \Tmax-0.4) 
    node[sloped,below] {$t^I_R$};
  \draw[->,thick]
    (\Zmax-0.2,\Tmax-0.4) 
      to[out=225,in=-225]        
    (\Zmax-0.2, -\Tmax+0.4) 
    node[sloped,above] {$t^{II}_R$};

  \draw[->, thick] (\Zmax+0.25,-\Tmax-0.25) -- (-\Zmax-0.25,\Tmax+0.25) node[above] {$u_G$};
  \draw[->, thick] (-\Zmax-0.25,-\Tmax-0.25) -- (\Zmax+0.25,\Tmax+0.25) node[above] {$v_G$};
 
  \node at (-1.5,0) {\circled{I}};
  \node at (1.5,0) {\circled{II}};
  \node at (0,-1.5) {WH};
  \node at (0,1.5) {BH};
  \node at (0,3.5) {Global Patch};
  \node at (0,-4) {};

\end{tikzpicture}
    \caption{}
  \end{subfigure}
  \caption{Three coordinate systems on AdS$_2$: (a) Global and Poincaré patches. (b) Global and Schwarzschild patches.}
  \label{Patches}
\end{figure}
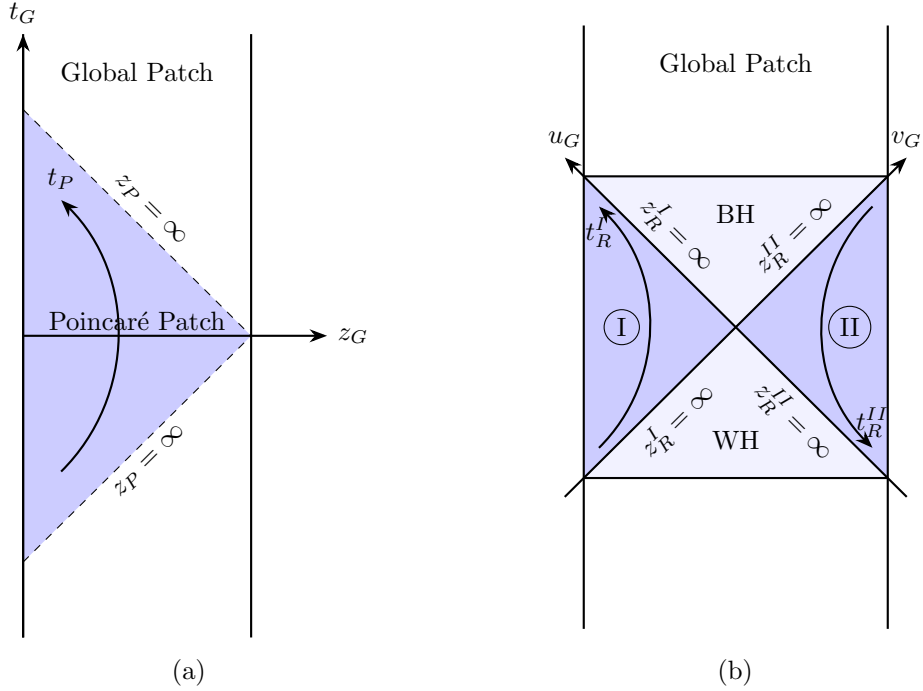

Furthermore, one can introduce Rindler (or equivalently Schwarzschild) coordinates to describe two black hole exterior patches in the global AdS$_2$ manifold:  

\begin{equation}
\label{Patch I}
\text{Patch \circled{I}}  
    \begin{cases}
  u_G = 2\tan^{-1}\left[\frac{\beta}{\pi}\tanh\left(\frac{\pi}{\beta}u_R^{I}\right)\right]-\frac{\pi}{2},  \\[4pt]
  v_G = 2\tan^{-1}\left[\frac{\beta}{\pi}\tanh\left(\frac{\pi}{\beta}v_R^{I}\right)\right]+\frac{\pi}{2},  \\
\end{cases}
\end{equation}
\begin{equation}
\label{Patch II}
\text{Patch \circled{II}}  
 \begin{cases}
  u_G = -2\tan^{-1}\left[\frac{\beta}{\pi}\tanh\left(\frac{\pi}{\beta}u_R^{II}\right)\right]+\frac{\pi}{2},  \\[4pt]
  v_G = -2\tan^{-1}\left[\frac{\beta}{\pi}\tanh\left(\frac{\pi}{\beta}v_R^{II}\right)\right]-\frac{\pi}{2}.  \\
\end{cases}
\end{equation}
The constant $\beta$ is related to the acceleration of an observer at fixed $z_R$ (or, equivalently to the ''black hole'' mass) and quantum mechanically corresponds to the inverse horizon temperature \cite{Unruh:1976db}. 

\noindent The resulting Rindler metric takes the form:

\begin{equation}
\label{Rindler coordinates}
    ds^2=\left(\frac{2\pi}{\beta}\right)^2\frac{1}{\sinh^2\left(\frac{2\pi}{\beta}z_R\right)}(-dt_R^2+dz_R^2)=-\left(\frac{2\pi}{\beta}\right)^2\frac{1}{\sinh^2(\frac{\pi}{\beta}(u_R-v_R))}du_Rdv_R,
\end{equation}

\noindent where $z_R\in(0,\infty)$, and $t_R\in\mathbb{R}$ in both patches, each of which is causally disconnected from the other by an event horizon at $z_R^{I},z_R^{II}=\infty$.  \\

Using the radial coordinate $r=r_h\coth(r_hz_R)$, one can also rewrite this metric in the Schwarzschild form:

\begin{equation}
\label{Schwarzschild}
    ds^2=-f(r)dt_R^2+\frac{dr^2}{f(r)},
\end{equation}

\noindent where $r \in (r_h,\infty)$, $f(r)=r^2-r_h^2$, and the horizon is now located at $r_h=\frac{2\pi}{\beta}$.\\

Each of the patches in Fig.\ref{Patches} (b) and eqs.\eqref{Patch I} and \eqref{Patch II} possesses its own asymptotic boundary and its own timelike Killing vector generating time evolution $t_R$ within that patch. The two exterior regions are connected through the interior of the eternal black hole geometry, forming a smooth but non-traversable wormhole in the sense of a connected AdS$_2$ spacetime with two asymptotic boundaries. In the holographic dual description, these boundaries correspond to two decoupled boundary quantum systems \cite{maldacena2001}. Quantum mechanically, the eternal geometry is dual to a thermofield-double (TFD) state of the two boundary theories \cite{maldacena2013}, whose entanglement purifies each boundary thermal density matrix and underlies the connected structure of the bulk spacetime \cite{ryu2006}.

\subsection{Schwarzian Dynamics}

Putting the bulk term of eq.(\ref{JT_Action}) on shell selects the background manifold and reduces the JT action to the GHY boundary term \cite{almheiri2015jt,maldacenastanford2016,engelsoy2016}. Performing the gravitational path integral is then equivalent to summing over AdS$_2$ manifolds with different boundaries. To study the dynamics of these boundaries we cut the manifold along a constant value of the dilaton in an arbitrary coordinate system. \\

Varying eq.(\ref{JT_Action}) with respect to the metric results in the vacuum equations of motion for the dilaton field:

\begin{equation}
    \label{Dilaton E.o.M}
\nabla_\mu \nabla_\nu \Phi - g_{\mu\nu}\nabla^2 \Phi + g_{\mu\nu}\Phi=0.
\end{equation}

\noindent Solving eq.(\ref{Dilaton E.o.M}) in Poincaré coordinates, eq.(\ref{Poincaré patch}), one obtains a divergent dilaton profile \cite{maldacenastanford2016}:

\begin{equation}
    \label{Dilaton profile}
    \Phi(t_P,z_P)=\frac{A+Bt_P+C(t_P^2+z_P^2)}{z_P},
\end{equation}

\noindent where $A$, $B$, and $C$ are arbitrary parameters of the theory. Clearly, the dilaton field diverges at the boundary $z_P=0$. However, the theory can be regulated by introducing a near-boundary cutoff \cite{engelsoy2016,maldacenastanford2016}.

The gravitational path integral is defined for metrics which are asymptotically AdS$_2$:

\begin{equation}
\label{Fefferman-Graham gauge}
    ds^2=\frac{1}{z^2}(-dt^2+dz^2) + \mathcal{O}(z^{-1}) \quad \text{as} \quad z\rightarrow0,
\end{equation}

\noindent  while imposing fixed divergent asymptotics for the dilaton field:

\begin{equation}
\label{Dilaton Asymptotics}
    \Phi=\frac{a}{2z} + \mathcal{O}(z) \quad \text{as} \quad z\rightarrow0,
\end{equation}

\noindent where $a$ is the regularised value of the dilaton. Performing an arbitrary coordinate transformation $u_P(u,v)$, $v_P(u,v)$, the first boundary condition, eq.\eqref{Fefferman-Graham gauge}, forces $\partial_{u}v_P=\partial_{v}u_P=0$ and $u_P(u,v)=v_P(u,v)$ at leading order in $z$ as $z\rightarrow0$:

\begin{equation}
    \begin{cases}
  u_P(u) = F(u)+\mathcal{O}(z),  \\[6pt]
  v_P(v) = F(v)+\mathcal{O}(z).
\end{cases}
\end{equation}

\noindent In the Poincaré patch, the boundary $z=\epsilon$ is therefore parameterized by:

\begin{equation}
    \begin{cases}
  t_P(t) = \frac{1}{2}(u_P+v_P) = F(t)+\mathcal{O}(\epsilon),  \\[6pt]
  z_P(t) = \frac{1}{2}(u_P-v_P) = \epsilon F'(t)+\mathcal{O}(\epsilon^2),
\end{cases}
\end{equation}

\noindent has induced metric $\sqrt{-h}=\frac{1}{\epsilon}+\mathcal{O}(\epsilon)$ and extrinsic curvature trace:

\begin{equation}
    \label{Extrinsic Curvature}
    K=1+\epsilon^2\left\{F(t),t\right\}+\mathcal{O}(\epsilon^4), 
\end{equation}

\noindent where the Schwarzian derivative is defined as:

\begin{equation}
    \{F,t\} := \frac{F'''(t)}{F'(t)} - \frac{3}{2} \left( \frac{F''(t)}{F'(t)} \right)^2.
\end{equation}

\noindent The second boundary condition, eq.\eqref{Dilaton Asymptotics}, forces the regulated value of the dilaton at this boundary to be $\Phi_{|\partial}=\frac{a}{2\epsilon}+\mathcal{O}(\epsilon)$.\\

Finally, the bulk gravitational action, eq.\eqref{JT_Action}, simplifies to the Schwarzian action governing the dynamics of the boundary:

\begin{equation}
    \label{Schwarzian_action}
    S_{\mathrm{JT}}[g,\Phi]
   = \frac{1}{8\pi G_N}
     \int_{\partial\mathcal{M}}\Phi_{|\partial}\,(K-1)\sqrt{-h}\,dt\,=\frac{a}{16\pi G_N}\int \left\{F(t),t\right\}dt
\end{equation}

\section{Scalar Correlators in JT Gravity}

We are interested in calculating the vacuum correlation of a massless scalar field $\phi$, including gravitational fluctuations. For simplicity, we assume a field decoupled from the dilaton, therefore ignoring backreaction. 

\subsection{Tree-Level Correlators}

The AdS$_2$ vacuum is fully specified by the two-point function $G_{12}=\langle\phi(x_1)\phi(x_2)\rangle$. In Lorentzian spacetimes there exist various Green functions; we focus on the Hadamard function:
\begin{equation}
   H_{12}=\langle\{\phi(x_1),\phi(x_2)\}\rangle, 
\end{equation}

\noindent which is related to the Feynman propagator $G_{12}^F=i\langle T\phi(x_1)\phi(x_2)\rangle$ by $H_{12}=2\operatorname{Im}G^F_{12}$.\\

The fixed background Hadamard function of the global AdS$_2$ vacuum is known to be \cite{spradlin1999vacuum}: 

\begin{equation}
\label{global correlator}
    H_{12} = -\frac{1}{2\pi}\ln \left| \frac{\cos(t_G -t_G')-\cos(z_G - z_G')}{\cos(t_G - t_G')+\cos(z_G + z_G')} \right|.
\end{equation}

\noindent One can express this function in Rindler coordinates, eq.(\ref{Patch I}) and eq.\eqref{Patch II}, to obtain a thermal vacuum in either of the two BH exteriors. To evaluate the correlator in only one patch, we choose the coordinates corresponding to that patch. For example in the left exterior:

\begin{equation}
\label{Hadamard LL}
    H_{12}^{LL} = -\frac{1}{2\pi} \ln \left| \frac{\sinh\frac{\pi}{\beta}(u^I_{R,1} - u_{R,2}^I)\sinh\frac{\pi}{\beta}(v_{R,1}^I - v_{R,2}^I)}{\sinh\frac{\pi}{\beta}(v_{R,1}^I - u_{R,2}^I)\sinh\frac{\pi}{\beta}(u_{R,1}^I - v_{R,2}^I)} \right|.
\end{equation}

Furthermore, one can also investigate the global vacuum correlation across both BH patches: 

\begin{equation}
\label{Hadamard LR}
    H_{12}^{LR} =  -\frac{1}{2\pi}\ln \left| \frac{\cosh\frac{\pi}{\beta}(u_{R,1}^I + v_{R,2}^{II})\cosh\frac{\pi}{\beta}(v_{R,1}^I + u_{R,2}^{II})}{\cosh\frac{\pi}{\beta}(v_{R,1}^I+ v_{R,2}^{II})\cosh\frac{\pi}{\beta}(u_{R,1}^I + u_{R,2}^{II}
    )} \right|.
\end{equation}

\subsection{Gravity Correlators}

In the Rindler patch the boundary reparametrization is given by $F(t_R)=\tanh\frac{\pi}{\beta}f(t_R)$ at leading order as $z_R\rightarrow 0$, where $f(t_R)$ is a function that satisfies:
\begin{equation}
\label{thermal disk Lorentzian}
    f(t_R+i\beta)=f(t_R)+i\beta, \quad f'(t_R) \ge 0.
\end{equation}

\noindent The first relation enforces that $f$ wraps around the Euclidean thermal circle once for each period $\beta$ in $\tau=it_R$. The second requirement guarantees $f$ is monotone and thus orientation preserving. These properties, after Wick rotation, identify $f$ as an element of the diffeomorphism group of the circle, $\mathrm{Diff}(\mathbb{S}^1)$, and justify interpreting it as a boundary time reparametrization.\\

Since the JT integral eq.(\ref{JT_Action}) always reduces to a boundary term after the background manifold is selected, integrating the Hadamard function over metrics is equivalent to an integration over the boundary function $f(t_R)$ \cite{maldacenastanford2016,engelsoy2016,saad2019jt,stanfordwitten2020,mertens2023review,blommaert2019}:

\begin{equation}
\label{Gravity Integral}
    \langle H_{12}\rangle=\operatorname{Re}\left(\frac{\int H_{12}[g]e^{iS_{JT}[g,\Phi]}\mathcal{D}g\mathcal{D}\Phi}{\int e^{iS_{JT}[g,\Phi]}\mathcal{D}g\mathcal{D}\Phi}\right)=\operatorname{Re}\left(\frac{\int H_{12}[f]e^{iS_{JT}[f]}d\mu[f]}{Z(\lambda)}\right),
\end{equation}

\noindent where $Z(\lambda)=\int e^{iS_{JT}[f]}d\mu[f]$ is the vacuum normalization, $H_{12}[f]$ is the bulk Hadamard function evaluated on the metric parametrized by $f$, and $d\mu[f]$ is the measure of integration over the symplectic manifold Diff$(\mathbb{S}^1)/SL(2,\mathbb{R})$ \cite{stanford2017localization}.

After Wick rotating the boundary action $iS_{JT}\rightarrow I$, rescaling the boundary circle to $\hat{\tau} = \frac{2\pi}{\beta}
 \tau$ and $\hat{f}(\hat{\tau})=\frac{2\pi}{\beta}f(\tau)$, and evaluating the Pfaffian measure $d\mu[\hat{f}]$ in terms of a Grassmann field $\psi$ \cite{stanford2017localization,saad2019jt}, the bulk quantum gravity correlator evaluates to:

\begin{equation}
\label{Pfaffian Euclidean functional integral}
     \langle H_{12}\rangle =  \operatorname{Re}\left(\frac{\int H_{12}[\hat{f}] e^{-\frac{1}{2} I[\hat{f},\psi]}\mathcal{D}\hat{f}\mathcal{D}\psi}{Z(\lambda)}\right),
\end{equation}

\noindent where the Euclidean action is a functional of the boundary functions $\hat{f}(\hat{\tau})$ and $\psi(\hat{\tau})$:

\begin{equation}
\label{Schwarzian Operator}
     I[\hat{f},\psi]=\int_0^{2\pi}\frac{\ddot{\hat{f}}^2}{\lambda^{2}\dot{\hat{f}}^2}-\frac{\dot{\hat{f}}^2}{\lambda^{2}} + \frac{\ddot{\psi}\dot{\psi}}{\dot{\hat{f}}^2}-\dot{\psi}\psi\ d\hat{\tau},
\end{equation}

\noindent with $\lambda^2=\frac{8G_N\beta}{a}$.\\

In the following we shall assume that the gravity coupling constant $\lambda\ll 1$ and fixed. This will justify a perturbative expansion in $\lambda$.

\subsection{Perturbative Expansion}

The saddle of the action eq.(\ref{Schwarzian_action}), subject to the periodicity condition eq.(\ref{thermal disk Lorentzian}), is $\hat{f}_s(\hat{\tau}) = \hat{\tau}$ \cite{maldacenastanford2016,engelsoy2016,mertens2023review}. Evaluated at this saddle, the two-sided correlators $\langle H_{12}\rangle$ exhibit pure exponential decay at late times, reproducing the familiar semiclassical behavior that underlies the tension with the unitarity of the dual boundary theory \cite{maldacena2001}. From the point of view of the gravitational path integral, this reflects the fact that $H_{12}(\hat f_s)\rightarrow 0$ at large times. To determine whether residual correlations survive, one must therefore examine fluctuations around the saddle and include subleading terms in the steepest-descent expansion in the parameter $\lambda$, following the approach of \cite{germani}. If subleading corrections generate non-decaying contributions to $\langle H_{12}\rangle$, then late-time correlations do not exponentially vanish despite the leading semiclassical decay. 

For small $\lambda$, one can use the method of stationary phase to asymptotically expand the functional part of the integral eq.(\ref{Pfaffian Euclidean functional integral}) in powers of $\lambda$ around the saddle point $\hat f_s$. In other words, we shall consider $\hat{f}(\hat{\tau}) = \hat{\tau} +\lambda \gamma(\hat{\tau})+\ldots$ at leading order in $\lambda$. By using the map of boundary correlators versus bulk correlators introduced in \cite{blommaert2019}, in \cite{germani} a linear ramp was found after an exponential decay for $\langle H_{12}^{LL}\rangle$ at quadratic order in $\lambda$. However, this behavior was in tension with the Schwarzian computation of \cite{Maldacena:2016upp}. Upon revisiting the calculation, we find that the discrepancy originates from a sign difference in the correlator $\langle\gamma(\hat\tau_1)\dot\gamma(\hat\tau_2)\rangle$ used in \cite{germani} (cf. \cite{qi2018}). Correcting this sign removes the linear growth in $\langle H_{12}^{LL}\rangle$ at order $\lambda^2$, as the relevant contributions cancel. The sign structure reflects the time-reversal properties of the derivative of the two-point function. As we will show, once this correction is implemented, the linear ramp instead appears in the cross-boundary correlator $H_{12}^{LR}$ connecting the two exteriors of the maximally extended geometry \eqref{Rindler coordinates}.\\

Expanding the action, eq.(\ref{Schwarzian Operator}), in powers of $\lambda$, we have:

\begin{equation}
\label{Schwarzian operator expanded}
    I[\hat{\tau}+\lambda\gamma(\hat{\tau}),\psi] = -\frac{2\pi}{\lambda^2} +I_0[\gamma,{\psi}]+\sum_{k=1}^\infty \lambda^{k}I_k[\gamma,\psi],
\end{equation}

\noindent where:

\begin{equation}
    I_0[\gamma,\psi] = \int_0^{2\pi}\ddot{\gamma}^2 - \dot{\gamma}^2 +\ddot{\psi}\dot{\psi} - \dot{\psi}\psi\,d\hat{\tau},
\end{equation}

\noindent and

\begin{equation}
    I_k[\gamma,\psi] = (-1)^k(k+1)\int_0^{2\pi}\dot{\gamma}^k (\ddot{\gamma}^2 +\ddot{\psi}\dot{\psi})\,d\hat{\tau}.
\end{equation}

\noindent Similarly, one can expand the bulk correlator $H_{12}[\hat{\tau}+\lambda\gamma]=\sum_0^{\infty}\lambda^{k}H_k[\gamma]$, to obtain the full perturbative expansion:

\begin{equation}
\label{Grav Cor Pert}
\begin{split}
\bigl\langle H_{12}\bigr\rangle =
& \operatorname{Re}\left(\frac{\int \Bigl(\sum_{k=0}^{\infty}\lambda^{k}H_k\Bigr)\,
  e^{\frac{\pi}{\lambda^2}}e^{-\frac{1}{2}I_0}\,e^{-\frac{1}{2}\sum_{n=1}^\infty \lambda^{n}I_n}\,\mathcal D\gamma}{Z_{1-loop}(\lambda)}\right)
  = \sum_{k=0}^\infty \lambda^{k}\langle\Gamma_k\bigr\rangle,\\[6pt]
\text{where:}\quad
&\Gamma_{0} = H_{0},\\
&\Gamma_{1} = \operatorname{Re} (H_{1} + H_{0}I_{1} ),\\
&\Gamma_{2} =  \operatorname{Re}\bigl(H_{2} + H_{1}I_{1} + \tfrac12\bigl(I_{1}^{2} - 3I_{2}\bigr)\bigr),\\
&\vdots
\end{split}
\end{equation}

    \noindent and where we used the fact that the Schwarzian partition function is one-loop exact \cite{stanford2017localization}, so that the normalization $Z(\lambda)$ is given exactly by $Z_{1-loop}(\lambda)$. In particular, although the partition function receives no higher-loop corrections, the correlator  itself is given by an infinite series in $\lambda$ arising from the numerator expansion. For each order in $\lambda$, $\langle \Gamma_k[\gamma, \psi] \rangle$ is a Gaussian integral involving correlations of the perturbation \cite{qi2018}:

\begin{equation}
\begin{aligned}
     & \langle T\gamma(\hat \tau_1)\gamma(\hat \tau_2)\rangle 
      = \int\gamma(\hat \tau_1)\gamma(\hat \tau_2)e^{-I_0[\gamma]}\mathcal{D}\gamma\\
     & = \frac{1}{2\pi}\left(1+\frac{\pi^2}{6}-\frac{1}{2}(\pi-\hat{\tau})^2-(\pi-\hat{\tau})\sin(\hat{\tau})+\frac{5}{2}\cos(\hat{\tau})\right),
\end{aligned}
\end{equation}

\noindent where $\hat{\tau} =|\hat{\tau}_1-\hat{\tau}_2|$. All odd orders of the expansion vanish and one is left with $\langle \Gamma_2\rangle$ as the first quantum correction.


\section{Results}

    To justify the perturbative expansion, we fix $\lambda^2=0.1$. Furthermore, we are interested in investigating the large time behavior of the gravitational path integral of the correlators in eq.\eqref{Hadamard LL} and eq.\eqref{Hadamard LR}. For this we fix a bulk point $z_{R,1}=z_{R,2}=z_R$, and set $t_{R,1}=0$ and $t_{R,2}=t_R$. Finally, in order to compare the $\beta$ dependence of the gravitationally dressed correlators in a geometrically well-defined way, we anchor the first bulk point at $t_P=0$ and $z_P=1$, and then find the corresponding $z_R$. For $\beta\leq\pi$, the chosen Rindler coordinate patch fails to cover the fixed Poincaré bulk point; we therefore restrict to $\beta>\pi$ for convenience. No physical singularity is associated with this $\beta=\pi$.

\subsection{Bulk Correlators}

\begin{figure}[htbp]
  \centering

  \begin{minipage}{0.48\textwidth}
    \centering
    \begin{tikzpicture}
      \node[anchor=south west, inner sep=0] (img) at (0,0)
        {\includegraphics[width=\linewidth]{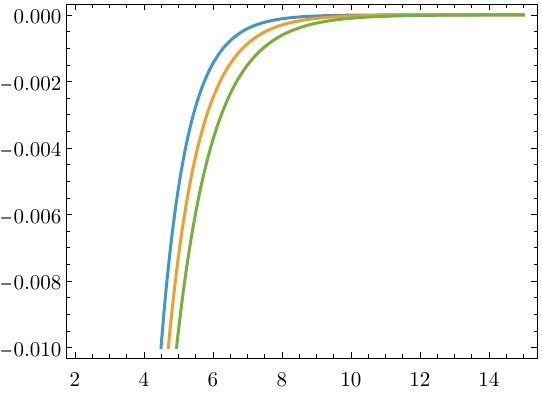}};
      \begin{scope}[x={(img.south east)}, y={(img.north west)}]
        \node at (1.02,0.05) {$t_R$};
        \node at (0.08,1.07) {$\langle \Gamma_{0}^{LL} \rangle$};
        \node at (0.8,0.33) {\plotlegend{$\beta=5$}{$\beta=6$}{$\beta=7$}};
      \end{scope}
    \end{tikzpicture}
    \captionof{figure}{One-sided bulk saddle correlator plotted for different values of $\beta$ and at $z_P=1$.}
    \label{H0LLBulkPlot}
  \end{minipage}\hfill
  \begin{minipage}{0.48\textwidth}
    \centering
    \begin{tikzpicture}
      \node[anchor=south west, inner sep=0] (img) at (0,0)
        {\includegraphics[width=\linewidth]{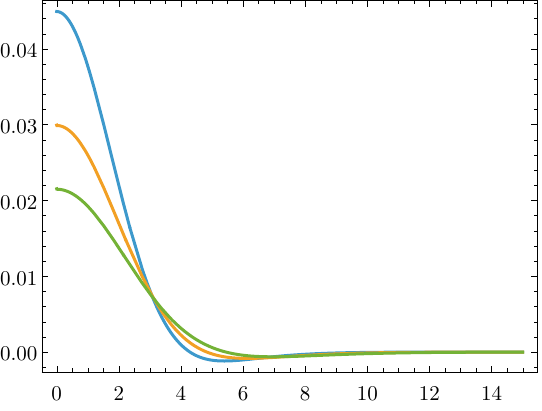}};
      \begin{scope}[x={(img.south east)}, y={(img.north west)}]
        \node at (1.02,0.05) {$t_R$};
        \node at (0.04,1.05) {$\langle \Gamma_{2}^{LL} \rangle$};
        \node at (0.8,0.78) {\plotlegend{$\beta=5$}{$\beta=6$}{$\beta=7$}};
      \end{scope}
    \end{tikzpicture}
    \captionof{figure}{First gravitational correction to the one-sided bulk correlator plotted for different values of $\beta$ and at $z_P=1$.}
    \label{H2LLBulkPlot}
  \end{minipage}
\end{figure}

First, let us investigate the one-sided bulk correlator. As described above, we set $t_{R,1}^I=0$, $t_{R,2}^I=t_R$, and $z_{R,1}^I=z_{R,2}^I=z_R=\frac{\beta}{\pi}\arctan\left(\frac{\pi}{\beta}\right)$. One finds that this correlator decays in time at both orders $\lambda^0$ and $\lambda^2$, as can be seen directly from Fig.\ref{H0LLBulkPlot} and Fig.\ref{H2LLBulkPlot}. This is in agreement with \cite{Maldacena:2016upp}. 

Specifically, taking the large time limit yields:

\begin{align}
\langle \Gamma_{0}^{LL} \rangle & \sim -\frac{2}{\pi}e^{-\frac{2\pi}{\beta}t_R}\sinh^2\left(\frac{2\pi}{\beta}z_R\right), \quad t_R\rightarrow \infty, \nonumber\\
\langle \Gamma_{2}^{LL} \rangle & \sim -\frac{8\pi}{\beta^2}t_R^2e^{-\frac{2\pi}{\beta}t_R}\sinh^2\left(\frac{2\pi}{\beta}z_R\right),\quad t_R\rightarrow\infty.
\end{align}

This behavior reflects the suppression of correlations between bulk points on the same side of the maximally extended geometry. The exponential decay is the standard semiclassical behavior of one-sided correlators in the eternal black hole geometry and is consistent with the analysis of \cite{Maldacena:2016upp}. In this approximation, the presence of the horizon gives rise to a thermal state, leading to late-time decay of one-sided correlators.

\begin{figure}[htbp]
  \centering

  \begin{minipage}{0.48\textwidth}
    \centering
    \begin{tikzpicture}
      \node[anchor=south west, inner sep=0] (img) at (0,0)
        {\includegraphics[width=\linewidth]{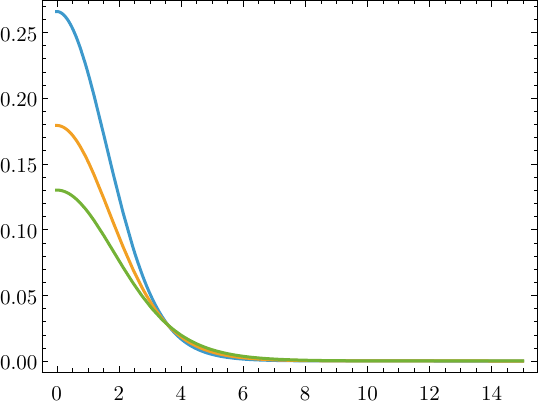}};
      \begin{scope}[x={(img.south east)}, y={(img.north west)}]
        \node at (1.02,0.05) {$t_R$};
        \node at (0.04,1.07) {$\langle \Gamma_{0}^{LR} \rangle$};
        \node at (0.8,0.78) {\plotlegend{$\beta=5$}{$\beta=6$}{$\beta=7$}};
      \end{scope}
    \end{tikzpicture}
    \captionof{figure}{Two-sided bulk saddle correlator plotted for different values of $\beta$ and at $z_P=1$.}
    \label{H0LRBulkPlot}
  \end{minipage}\hfill
  \begin{minipage}{0.48\textwidth}
    \centering
    \begin{tikzpicture}
      \node[anchor=south west, inner sep=0] (img) at (0,0)
        {\includegraphics[width=\linewidth]{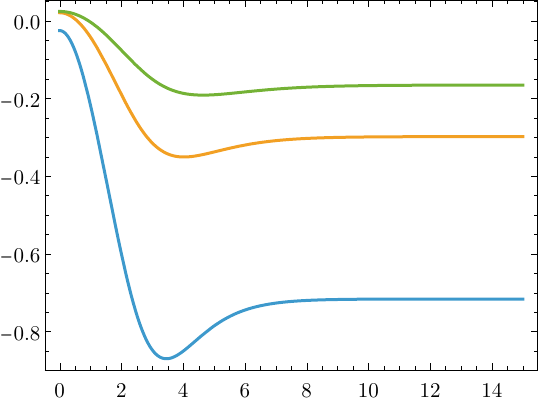}};
      \begin{scope}[x={(img.south east)}, y={(img.north west)}]
        \node at (1.02,0.05) {$t_R$};
        \node at (0.08,1.08) {$\langle \Gamma_{2}^{LR} \rangle$};
        \node at (0.8,0.465) {\plotlegend{$\beta=5$}{$\beta=6$}{$\beta=7$}};
      \end{scope}
    \end{tikzpicture}
    \captionof{figure}{First gravitational correction to the two-sided bulk correlator plotted for different values of $\beta$ and at $z_P=1$.}
    \label{H2LRBulkPlot}
  \end{minipage}

\end{figure}

In contrast, the two-sided correlator develops a qualitatively different late-time structure. Setting $t_{R,1}^I=0$, $t_{R,2}^{II}=t_R$, and $z_{R,1}^I=z_{R,2}^{II}=z_R=\frac{\beta}{\pi}\arctan\left(\frac{\pi}{\beta}\right)$, the bulk two-sided correlator decays to zero for large time at tree level only, as shown in Fig.\ref{H0LRBulkPlot}. At order $\lambda^2$ the correlator asymptotes to a negative but constant value, as seen in Fig.\ref{H2LRBulkPlot}.

Taking the large time limit, one obtains:

\begin{align}
\label{Asymptote LR}
\langle \Gamma_{0}^{LR} \rangle & \sim \frac{2}{\pi}e^{-\frac{2\pi}{\beta}t_R}\sinh^2\left(\frac{2\pi}{\beta}z_R\right), \quad t_R\rightarrow \infty, \nonumber \\
\langle \Gamma_{2}^{LR} \rangle & \sim -\frac{2z_R}{\pi\beta}\cosh{\left(\frac{2\pi}{\beta}z_R\right)}\sinh{\left(\frac{2\pi}{\beta}z_R\right)},\quad t_R\rightarrow\infty.
\end{align}

\noindent Although the leading saddle contribution $\langle \Gamma_{0}^{LR} \rangle$ decays exponentially with time (Eq.~\eqref{Asymptote LR} and Fig.~\ref{H0LRBulkPlot}), the next-to-leading correction $\langle \Gamma_2^{LR}\rangle$ alters the late-time behavior, generating a linear ramp that drives the correlator toward a constant, power-suppressed plateau (Fig.~\ref{H2LRBulkPlot}). Thus, non-vanishing correlations across the two exterior patches persist within the perturbative expansion.

\begin{figure}[htbp]
  \centering

  \begin{minipage}{0.48\textwidth}
    \centering
    \begin{tikzpicture}
      \node[anchor=south west, inner sep=0] (img) at (0,0)
        {\includegraphics[width=\linewidth]{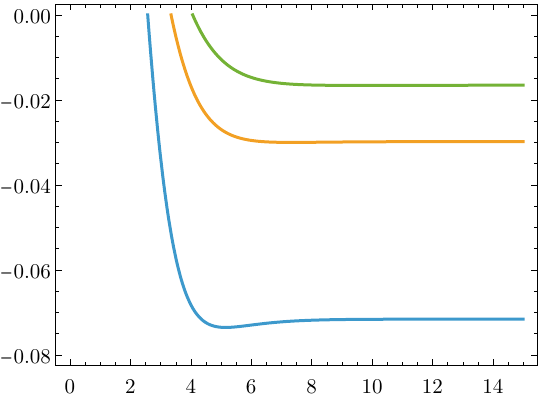}};
      \begin{scope}[x={(img.south east)}, y={(img.north west)}]
        \node at (1.02,0.05) {$t_R$};
        \node at (0.08,1.07) {$\langle H_{12}^{LR} \rangle$};
        \node at (0.8,0.43) {\plotlegend{$\beta=5$}{$\beta=6$}{$\beta=7$}};
      \end{scope}
    \end{tikzpicture}
    \captionof{figure}{Two-sided bulk gravity correlator plotted for different values of $\beta$, $\lambda^2=0.1$, and at $z_P=1$.}
    \label{O2LRBulkbetaPlot}
  \end{minipage}\hfill
  \begin{minipage}{0.48\textwidth}
    \centering
    \begin{tikzpicture}
      \node[anchor=south west, inner sep=0] (img) at (0,0)
        {\includegraphics[width=\linewidth]{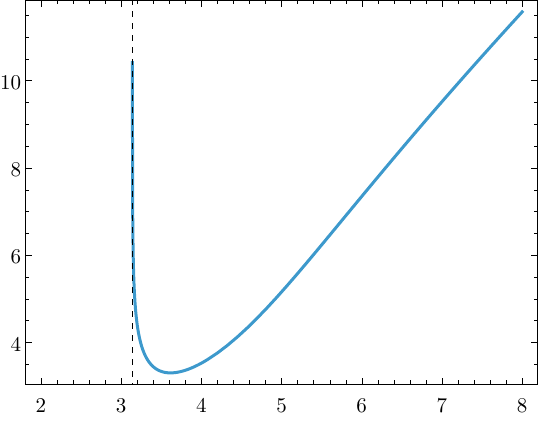}};
      \begin{scope}[x={(img.south east)}, y={(img.north west)}]
        \node at (1.02,0.05) {$\beta$};
        \node at (0.04,1.05) {$t_{dip}$};
      \end{scope}
    \end{tikzpicture}
    \captionof{figure}{Minimum of the two-sided bulk gravity correlator plotted as a function of $\beta$,  for $\lambda^2=0.1$, and at $z_P=1$.}
    \label{tdipbeta}
  \end{minipage}

\end{figure}

Combining both contributions, the bulk two sided gravity correlator 
$\langle H_{12}^{LR}\rangle \sim \langle \Gamma_0^{LR}\rangle+\lambda^2\langle \Gamma_2^{LR}\rangle$
exhibits a dip–ramp–plateau structure, as shown in Fig.~\ref{O2LRBulkbetaPlot}. In this framework, a ramp and plateau arise already at order $\lambda^2$, providing a perturbative realization of the dip–ramp–plateau behavior entirely within the disk topology, without invoking additional topologies or genuinely nonperturbative $e^{-S}$ effects.\\

Finally, defining $t_{\rm dip}$ as the time at which $\langle H_{12}^{LR} \rangle$ reaches its minimum, we find that $t_{\rm dip}$ scales linearly with $\beta$ when the fixed bulk point lies well within the Rindler patch, as shown in Fig.~\ref{tdipbeta}. Linear dependence of the dip time on the inverse temperature is likewise encountered in nonperturbative analyses of spectral observables \cite{mertens2023review}, although in that context additional entropy-dependent scales also enter. The same plot illustrates the increasing redshift of the Rindler time coordinate $t_R$ as the fixed bulk point approaches the boundary of the chosen Rindler patch near $\beta=\pi$.

\subsection{Boundary Correlators: Relation to AdS/CFT}

\noindent To obtain the boundary CFT correlators, in AdS/CFT one takes the near-boundary limit of the bulk two-point functions. For a massless scalar field in AdS$_2$, the scaling dimension is $\Delta = 1$; hence the boundary correlators are extracted by taking the limit \cite{Witten:1998qj}:

\begin{equation}
   \langle \tilde{H}_{12}\rangle = \lim_{z\rightarrow0}(z^{-2}\langle H_{12}\rangle).  
\end{equation}

\noindent Applying this to eq.\eqref{Hadamard LL} and eq.\eqref{Hadamard LR} results in the boundary saddle CFT correlators:

\begin{align}
\label{boundary saddle}
 \tilde{H}_{12}^{LL}  & = -\frac{2\pi}{\beta^2}\frac{1}{\sinh^2{(\frac{\pi}{\beta}(t_{R,1}^I-t_{R,2}^I))}},\nonumber \\
\tilde{H}_{12}^{LR}  & = \frac{2\pi}{\beta^2}\frac{1}{\cosh^2{(\frac{\pi}{\beta}(t_{R,1}^I+t_{R,2}^{II}))}}.
\end{align}\\
 
 One expects the boundary correlators to retain the features of their corresponding bulk functions. Setting $t_{R,1}^I=0$ and $t_{R,2}^I=t_{R,2}^{II}=t_R$, and taking the large time limit yields:

\begin{align}
\tilde{H}_{12}^{LL} &\sim -\frac{8\pi}{\beta^2} e^{-\frac{2\pi}{\beta} t_R}, 
&\quad
\tilde{H}_{12}^{LR} &\sim \frac{8\pi}{\beta^2} e^{-\frac{2\pi}{\beta} t_R},
\qquad &t_R \to \infty, \label{eq:H12_asymp}\\
\langle \tilde{\Gamma}_{2}^{LL} \rangle &\sim -\frac{32\pi^3}{\beta^4} t_R^2 e^{-\frac{2\pi}{\beta} t_R},
&\quad
\langle \tilde{\Gamma}_{2}^{LR} \rangle &\sim -\frac{4}{\beta^2},
\qquad &t_R \to \infty. \label{eq:Gamma2_asymp}
\end{align}
 
\noindent Clearly, the one-sided correlator decays exponentially for large time at both orders $\lambda^0$ and $\lambda^2$. However, the cross-boundary correlator decays exponentially at order $\lambda^0$ only. Furthermore, one observes that the height of the plateau goes to zero as $\beta\rightarrow\infty$, in accordance with \cite{mertens2017}. \\

\begin{figure}[htbp]
  \centering

  \begin{minipage}{0.48\textwidth}
    \centering
    \begin{tikzpicture}
      \node[anchor=south west, inner sep=0] (img) at (0,0)
        {\includegraphics[width=\linewidth]{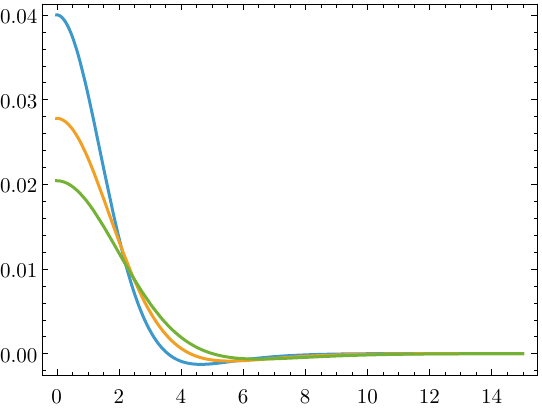}};
      \begin{scope}[x={(img.south east)}, y={(img.north west)}]
        \node at (1.02,0.05) {$t_R$};
        \node at (0.08,1.07) {$\langle \tilde{\Gamma}_{2}^{LL} \rangle$};
        \node at (0.8,0.77) {\plotlegend{$\beta=5$}{$\beta=6$}{$\beta=7$}};
      \end{scope}
    \end{tikzpicture}
    \captionof{figure}{First gravitational correction to the one-sided boundary correlator plotted for different values of $\beta$.}
    \label{H2LLBdyPlot}
  \end{minipage}\hfill
  \begin{minipage}{0.48\textwidth}
    \centering
    \begin{tikzpicture}
      \node[anchor=south west, inner sep=0] (img) at (0,0)
        {\includegraphics[width=\linewidth]{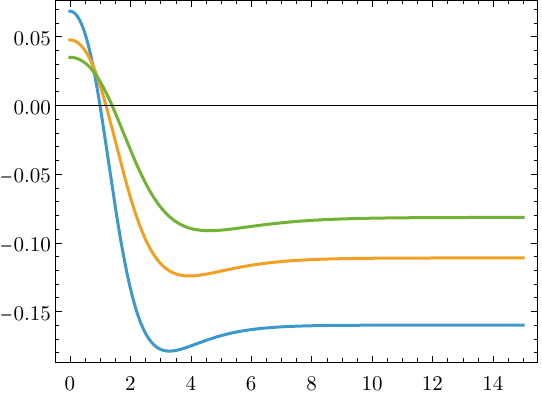}};
      \begin{scope}[x={(img.south east)}, y={(img.north west)}]
        \node at (1.02,0.05) {$t_R$};
        \node at (0.08,1.07) {$\langle \tilde{\Gamma}_{2}^{LR} \rangle$};
        \node at (0.8,0.77) {\plotlegend{$\beta=5$}{$\beta=6$}{$\beta=7$}};
      \end{scope}
    \end{tikzpicture}
    \captionof{figure}{First gravitational correction to the cross-boundary correlator plotted for different values of $\beta$.}
    \label{H2LRBdyPlot}
  \end{minipage}

\end{figure}

Plotting the full expression for the next to leading order of both correlators, one obtains the same dip-ramp-plateau features, shown in Fig.\ref{H2LLBdyPlot} and Fig.\ref{H2LRBdyPlot}. The crucial difference is that the height of the plateau of the one-sided correlator is zero for all values of $\beta$, whereas the height of the plateau of the cross-boundary correlator is $\beta$ dependent.

\begin{figure}[htbp]
  \centering

  \begin{minipage}{0.48\textwidth}
    \centering
    \begin{tikzpicture}
      \node[anchor=south west, inner sep=0] (img) at (0,0)
        {\includegraphics[width=\linewidth]{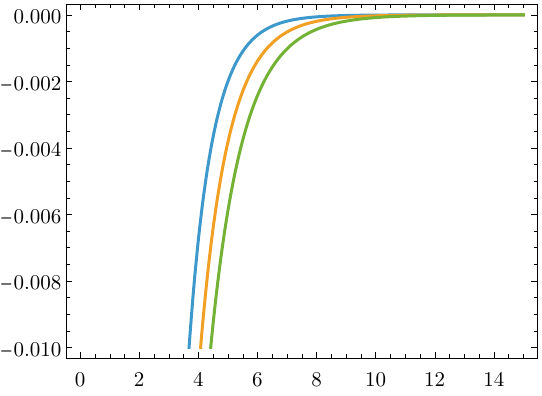}};
      \begin{scope}[x={(img.south east)}, y={(img.north west)}]
        \node at (1.02,0.05) {$t_R$};
        \node at (0.1,1.07) {$\langle \tilde{H}_{12}^{LL} \rangle$};
        \node at (0.8,0.33) {\plotlegend{$\beta=5$}{$\beta=6$}{$\beta=7$}};
      \end{scope}
    \end{tikzpicture}
    \captionof{figure}{One-sided boundary gravity correlator plotted for different values of $\beta$.}
    \label{O2LLBdyPlot}
  \end{minipage}\hfill
  \begin{minipage}{0.48\textwidth}
    \centering
    \begin{tikzpicture}
      \node[anchor=south west, inner sep=0] (img) at (0,0)
        {\includegraphics[width=\linewidth]{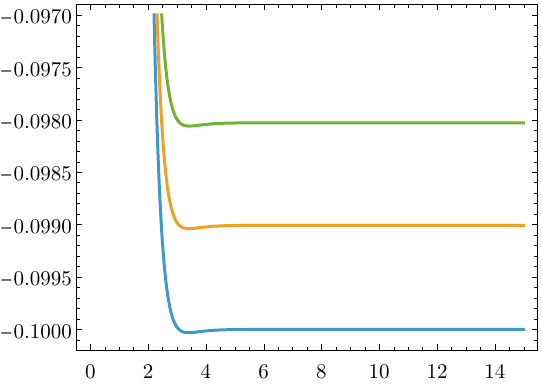}};
      \begin{scope}[x={(img.south east)}, y={(img.north west)}]
        \node at (1.02,0.05) {$t_R$};
        \node at (0.07,1.05) {$\langle \tilde{H}_{12}^{LR} \rangle$};
        \node at (0.8,0.77) {\plotlegend{$\beta=2.00$}{$\beta=2.01$}{$\beta=2.02$}};
      \end{scope}
    \end{tikzpicture}
    \captionof{figure}{Cross-boundary gravity correlator plotted for different values of $\beta$.}
    \label{O2LRBdyPlot}
  \end{minipage}

\end{figure}

On the other hand, plotting both orders of the boundary gravity correlators, one clearly observes the complete dip–ramp–plateau structure emerging only for the cross-boundary correlator. As shown in Fig.\ref{O2LLBdyPlot}, the one-sided boundary correlator decays exponentially at all temperatures, reflecting the loss of correlations across the horizon and reproducing the expected semiclassical behavior. In contrast, Fig.\ref{O2LRBdyPlot} shows the cross-boundary correlator displaying a clear dip–ramp–plateau structure whose amplitude and timescale depend on~$\beta$. The initial dip corresponds to the semiclassical decay, while the subsequent ramp and saturation indicate the re-emergence of correlations between the two boundaries.

Within the perturbative framework, this late-time plateau provides a realization of the qualitative dip–ramp–plateau pattern familiar from studies of spectral observables in SYK \cite{saad2018semiramp,cotler2017blackholes}, however, here arising entirely from fluctuations around the principal saddle rather than from genuinely nonperturbative effects.\\

\begin{figure}[htbp]
  \centering

  \resizebox{0.6\textwidth}{!}{%
    \begin{tikzpicture}
      \node[anchor=south west, inner sep=0] (img) at (0,0)
        {\includegraphics[width=\linewidth]{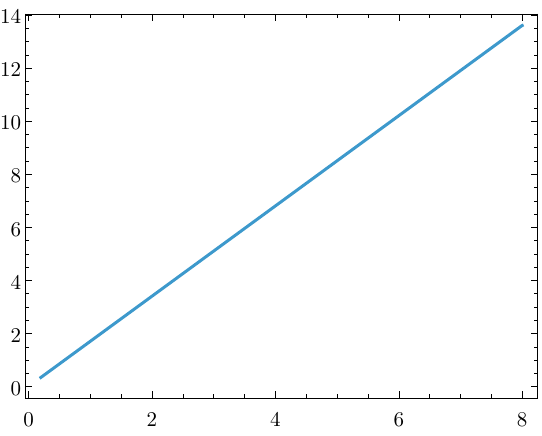}};

      \begin{scope}[x={(img.south east)}, y={(img.north west)}]
        \node[font=\Huge] at (1.03,0.03) {$\beta$};
        \node[font=\Huge] at (0.0,1.07) {$t_{dip}$};
      \end{scope}
    \end{tikzpicture}
  }

  \caption{Minimum of the cross-boundary gravity correlator plotted as a function of $\beta$ and  for $\lambda^2=0.1$.}
  \label{tdipbdy}
\end{figure}

Finally, defining $t_{\rm dip}$ as the time at which $\langle \tilde{H}_{12}^{LR} \rangle$ reaches its minimum, we find that $t_{\rm dip}$ scales linearly with $\beta$, as shown in Fig.~\ref{tdipbdy}. Linear dependence of the dip time on the inverse temperature is likewise encountered in nonperturbative analyses of spectral observables \cite{mertens2023review}, although in that context additional entropy-dependent scales also enter.

\section{Perturbative vs Non-Perturbative}
It is useful to clarify the relation between the perturbative and nonperturbative regimes in the language of the gravitational path integral. Schematically, the partition function admits a saddle-point expansion of the type:
\[
Z \sim e^{-S_{\mathrm{dom}}/\lambda^2}
\sum_{n=0}^{\infty} a_n \lambda^n
\;+\;
\sum_{i} e^{-S_i/\lambda^2}
\sum_{n=0}^{\infty} b_{n,i} \lambda^n ,
\]
where $S_{\mathrm{dom}}$ denotes the action of the dominant saddle (the disk geometry in our case), and the $S_i$ correspond to subleading saddles exponentially suppressed in $1/\lambda^2$ capturing genuinely nonperturbative contributions, e.g. higher-topology configurations.

In our analysis, the dip–ramp–plateau structure arises entirely from the perturbative series around the dominant disk saddle. The entropy-suppressed effects associated with additional saddles are not required to generate the ramp and plateau at the order we considered. This does not exclude the eventual relevance of nonperturbative contributions at parametrically late times; rather, it shows that qualitatively nontrivial late-time structure is already encoded in the perturbative sector of the theory.

From the perspective of the dual description, this separation between power-suppressed and entropy-suppressed contributions maps directly onto the large-$N$ expansion of the partition function. In the SYK/JT correspondence, the gravitational coupling in the bulk scales as $\lambda^2 \sim G_N \sim 1/N$ in the large-$N$ limit. For the particular cross-boundary observable we study, the natural dual is the two-sided correlator in a pair of maximally entangled SYK systems prepared in the thermofield-double state. In this setting, the plateau we find corresponds to a contribution of order $1/N$ in the boundary theory, whereas entropy-controlled nonperturbative effects are expected to scale as $e^{-S} \sim e^{-N}$ \cite{cotler2017blackholes}. The cross-boundary plateau obtained from the disk saddle is thus parametrically larger than genuinely nonperturbative contributions, and therefore dominates the late-time behavior within the regime of validity of the expansion.

More broadly, our results suggest that the purely semiclassical picture, characterized by exponential decay of all correlators in a fixed black hole background, is too naive. Once subleading corrections in the perturbative expansion are included, qualitatively new late-time structures can emerge. This indicates that essential ingredients relevant to the information paradox may already be encoded in controlled quantum fluctuations around a single saddle geometry \cite{germani}. It would be interesting to compute the cross-boundary two-point function in a pair of maximally entangled SYK systems prepared in the thermofield-double state, beyond leading order in $1/N$, to be able to get a direct test of the perturbative mechanism identified here in the context of AdS/CFT. We leave this interesting question for future work.

\section{Conclusion}

In this work we have shown that a dip–ramp–plateau structure emerges perturbatively from bulk correlators in Jackiw–Teitelboim gravity within the disk topology, without invoking topology change or genuinely nonperturbative saddles. By evaluating the gravitationally dressed Hadamard function through the next-to-leading term in the steepest-descent expansion of the Schwarzian path integral, we found that correlators evaluated in a single black hole exterior region decay exponentially at all orders considered, reproducing the semiclassical suppression. In contrast, correlators linking the two exteriors display a qualitatively different behavior: after the initial semiclassical decay, the $\mathcal O(\lambda^2)$ correction generates a linear ramp followed by saturation to a constant value. At this perturbative order, both bulk and boundary observables exhibit the complete dip–ramp–plateau pattern. Higher orders in $\lambda^2$ eventually lead to exponential growth, indicating that the perturbative series is asymptotic. This is similar to the case of the same side correlators studied in \cite{jacopo}. The implications of this asymptotic structure are left for future research.

\acknowledgments

The authors wish to thank Jorge Russo for discussions, Roberto Emparan for comments on the first draft of the paper and Laia Montell\`a for earlier discussions. MK is supported by the Joan Or\'o fellowship. Our research is financed by the
grant PID2022-136224NB-C22, funded by MCIN\allowbreak/\allowbreak AEI\allowbreak/10.13039\allowbreak/501100011033\allowbreak/\allowbreak FEDER,
UE, by the grant 2021-SGR00872 and CEX2024-001451-M funded by MICIU/AEI/10.13039/501100011033.

\bibliographystyle{JHEP} 
\bibliography{references}
\end{document}